\documentstyle[prd,aps,epsfig,preprint,tighten]{revtex}
\begin{document}
\preprint{                                                BARI-TH/299-98}
\draft
\title{     Vacuum oscillations and variations of solar neutrino rates\\ 
	             in SuperKamiokande and Borexino}
\author{      	B.~Fa\"{\i}d,$^a$  
		G.~L.~Fogli,$^b$ 
		E.~Lisi,$^b$ and 
		D.~Montanino~$^b$			}
\address{
 $^a$Institut de Physique, Universit{\'e} des Sciences et de la Technologie,
     DZ-16111 Algiers, Algeria\\
 $^b$Dipartimento di Fisica and Sezione INFN di Bari, 
     Via Amendola 173, I-70126 Bari, Italy		}
\maketitle
\begin{abstract}
The vacuum oscillation solution to the solar neutrino problem predicts
characteristics variations of the observable neutrinos rates, as a result
of the $L/E_\nu$ dependence of the $\nu_e$ survival probability
($L$ and $E_\nu$ being the neutrino pathlength and energy, respectively).
The $E_\nu$-dependence can be studied through distortions of the recoil
electron spectrum in the SuperKamiokande experiment. The $L$-dependence
can be investigated through a Fourier analysis of the signal in 
the SuperKamiokande and Borexino experiments. We discuss in detail the
interplay among such observable variations of the signal,
and show how they can help to test and constrain the vacuum oscillation
solution(s). The analysis includes the 374-day SuperKamiokande data.
\end{abstract}

\section{Introduction}

	Neutrino flavor oscillations \cite{Pont} with wavelength comparable 
to the Earth-Sun distance \cite{Gl87} represent a  solution \cite{Kr96} to 
the deficit of solar $\nu$'s \cite{Ba89} observed in the four pioneering 
underground experiments Homestake \cite{Da94}, Kamiokande \cite{Fu96}, SAGE 
\cite{Ab96}, and GALLEX \cite{Ha96}, as compared to the standard solar model 
predictions \cite{BP95}. The recent SuperKamiokande data \cite{To97} 
confirm the deficit, and can be interpreted within the vacuum oscillation 
hypothesis as well \cite{Fo97,Ha97}. The planned Borexino solar neutrino 
experiment \cite{Prop,Bwww} (in construction), designed to detect 
monochromatic $^7$Be neutrinos ($E_\nu =0.86$ MeV), is expected to 
test this hypothesis with unprecedented sensitivity \cite{Be95,Ra95,Ba97}.

	An update of the current neutrino flux measurements 
\cite{Fu96,To97,La97,Ga97,Hp97} is given in Table~I. Figure~1 shows our 
vacuum oscillation fit to the data of Table~I, as obtained from a $\chi^2$ 
analysis (including solar model uncertainties as in \cite{Fo95}). We have 
assumed, for simplicity, two neutrino families. It can be seen that four 
regions (A, B, C, and D) are allowed at 95\% C.L., the absolute minimum 
being located within the solution B ($\chi^2_{\rm min}=3.4$ and 
$N_{\rm DF}=3=5-2$).

	Various tests can be envisaged to discriminate among the four 
solutions in Fig.~1. In particular, since the neutrino oscillation length is 
proportional to the pathlength-to-energy ratio $L/E_\nu$, deviations of event 
distributions from the expected shape in either $L$ or $E_\nu$ (or related 
parameters) represent direct tests of neutrino vacuum oscillations
(see, e.g., \cite{Kr97,Faid,Mi97}). It turns out that, in general, 
SuperKamiokande is more (less) sensitive than Borexino to  $E_\nu$-related 
($L$-related) spectral shape deviations; therefore, the two experiments 
provide complementary tools to study the vacuum oscillation hypothesis. 
The purpose of this work is to investigate in detail the tests of the 
vacuum oscillation hypothesis that can be performed at SuperKamiokande and 
Borexino, and their interplay. The results will be shown in a form that 
makes easy to derive the experimental accuracy needed to perform a 
specific test.

 	The plan of this paper is as follows. In Sec.~II we discuss
the tests of energy spectra deviations, in the light of the recent 
SuperKamiokande results. In Sec.~III we discuss in detail the Fourier 
analysis of the signal, building upon our previous work \cite{Four}.
In Sec.~III we apply these tests to SuperKamiokande and Borexino, both 
separately and jointly. In Sec.~IV we draw our conclusions. Some technical 
aspects of our analysis are described in the Appendix.

\section{$E_\nu$-related tests: Average electron kinetic energy}

	Both SuperKamiokande and Borexino can measure the energy spectrum of
recoil electrons from neutrino scattering. The standard (i.e., no oscillation) 
SuperKamiokande spectrum can be found in Fig.~4 of Ref.~\cite{Fo97}. For 
completeness, we show in Fig.~2 the standard  electron energy spectrum in 
Borexino (details about the inclusion of energy threshold and resolution 
effects are given in the Appendix). The main contribution to the spectrum 
in Fig.~2 is given by the 0.86 MeV $^7$Be line, which is responsible for the 
Compton edge at $\sim 0.66$ MeV. The edge is smeared by the finite energy 
resolution. The rise at low energies is due to $pp$ neutrinos, while the 
tail at high energies is basically due to CNO neutrinos. The prospective 
analysis window is also indicated in Fig.~2.  The standard neutrino
fluxes have been taken from \cite{BP95}.

	Distortions of the neutrino energy spectrum due to oscillations
are generally reflected (although somewhat degraded) in the electron energy 
spectrum. An effective parametrization of such distortions is given by the 
fractional variation of the average kinetic energy $\langle T \rangle$ of 
the electron, an approach extensively developed  in \cite{Kr97} and applied 
to the SuperKamiokande data in \cite{Fo97,Fo98}. In particular, the analysis 
\cite{Fo98} of the most recent (374 day) measurements of the electron 
spectrum in SuperKamiokande \cite{To97,Na98,In97,Su97,It97,So98} gives a 
fractional deviation
\begin{equation}
\frac{\Delta\langle T\rangle}{\langle T\rangle}\times 100
 = 0.95 \pm 0.73 \ .
\label{DT}
\end{equation}
Such deviation is consistent with zero (no oscillation) at the level
of $1.2\sigma$, and disfavors scenarios predicting negative values
for $\Delta \langle T\rangle/{\langle T\rangle}$.

	Concerning Borexino, the expected {\em shape\/} variations of the 
energy distribution are very small, since the main contribution to the 
electron spectrum comes from a monoenergetic source of neutrinos ($^7$Be),
rather than from a continuous source as in SuperKamiokande ($^8$B).
Neutrino oscillations are expected to change significantly
the global Borexino rate but not its energy spectrum.

\section{$L$-related tests: Fourier expansion of the signal}

	The variations of the solar neutrino pathlength $L$ due to the 
eccentricity of the Earth orbit produce a geometrical $(1/L^2)$ modulation of
the neutrino flux. Additional semiannual modulations are expected in the
presence of neutrino oscillations \cite{Pome}. The Fourier analysis
of the measured flux represents an effective tool to study both kinds of
$L$-related modulations \cite{Four}.

	In this Section we outline the Fourier analysis of the signal 
observable in solar neutrino experiments. In the first three subsections,
we describe the general formalism and the results for the ``no oscillation''
and ``$2\nu$ oscillation'' cases (see also Ref.~\cite{Four} for further 
details). In the last two subsections we generalize 
the analysis to $3\nu$ oscillations and then discuss a useful check of 
both the symmetry properties and the estimated uncertainties of the signal.

\subsection{General formalism}

	The Earth orbit radius, $L$, varies periodically in time $(t)$ around 
its average value, $L_0=1.496\times 10^8$ km, according to
\begin{equation}
L(t) = L_0\left(1-\varepsilon\, \cos\frac{2\,\pi\,t}{T}\right)
+O(\varepsilon^2)\ ,
\label{Lt}
\end{equation}
where $\varepsilon=0.0167$ is the orbit eccentricity and $T=1$~yr
($t=0$ at the perihelion). Terms of $O(\varepsilon^2)$
or higher are negligible for our purposes.

	The neutrino signal $S$ is also, in general, a periodic function.
Assuming a constant background $B$, the total observed neutrino rate $R$ is:
\begin{equation}
	R(t)	=	B+S(t)\ .
\label{Rt}
\end{equation}
For symmetry reasons, the analysis can be restricted to the time interval
$[0,T/2]$.%
\footnote{Experimental tests of the symmetry properties of the signal 
are discussed at the end of this section.}
It is understood that events collected in subsequent half-years must be 
symmetrically folded in this interval. The data sample consists then of $N$ 
events collected at different times $\{ t_i\}_{1\leq i\leq N}$, with 
$t_i\in [0,\,T/2]$ and $N$ equal to the total sum of background and signal
events, $N=N_S+N_B$. Notice that, in general, one can determine $N_B$ and 
$N_S$ but cannot distinguish background and signal on an event-by-event basis.

	The expansion of the signal in terms of Fourier components $f_n$ reads
\begin{equation}
	S(t)	=	S\left( 1+2\sum_{n=1}^{\infty}
			f_n\cos\frac{2\pi n t}{T}\right)\ ,
\label{St}
\end{equation}
where $S$ is the time-averaged signal
\begin{equation}
	S	=	\frac{2}{T}\int^{T/2}_0\! dt\,S(t)\ . 
\label{S}
\end{equation}
The $n$-th harmonic corresponds to a period of $1/n$~yr. The explicit form  
of $f_n$ reads
\begin{eqnarray}
	f_n 	& = & 	\frac{2}{ST} \int^{T/2}_0\! dt\,
			R(t) \cos\frac{2\pi n t}{T}
\label{fntheo}\\
 		& = & 	\frac{1}{N_S}\sum_{i=1}^{N_S+N_B}
 			\cos\frac{2\pi n t_i}{T}\ \ \ \ \ \
 			(0\leq t_i\leq T/2)\ ,
\label{fnexpt}
\end{eqnarray}
where Eqs.~(\ref{fntheo}) and (\ref{fnexpt}) represent the theoretical
definition and the experimental determination of the $f_n$'s, respectively
\cite{Four}.

	Assuming purely statistical fluctuations of the signal and of the 
background, the variance of the $f_n$'s  reads:
\begin{equation}
	{\rm var}(f_n)	= \frac{1+f_{2n}+N_B/N_S}{2\,N_S}\ .
\label{varfn}
\end{equation}
It turns out that the values of the $f_{2n}$'s are $\ll 1$ in all cases
of practical interest. Therefore, to a good approximation, the one-sigma  
statistical 
error $\sigma_f=\sqrt{{\rm var}(f_n)}$  affecting $f_n$ is given by
\begin{equation}
	\sigma_f \simeq \sqrt{\frac{N_S+N_B}{2\,N^2_S}}
\label{sigma}
\end{equation}
for {\em any\/} $n$. The correlations between the statistical errors
of different harmonics are also negligible \cite{Four}.

	Finally, the general expression of the signal $S$ expected in the 
presence of oscillations is 
\begin{equation}
S^{\rm osc}(t) \propto \frac{L_0^2}{L^2(t)}\int\! dE\, \lambda(E) 
			\left[ \sigma_e(E) P(E,t)+ \sigma_x(E) 
			(1-P(E,t))\right],
\label{Sosc}
\end{equation}
where $E$ is the neutrino energy, $\lambda$ is the neutrino energy
spectrum, $\sigma_e$ ($\sigma_x$) is the $\nu_e$  ($\nu_x$, $x=\mu,\,\tau$) 
interaction cross section, and $P$ is the $\nu_e$ survival probability, 
which varies in time through $L(t)$ \cite{Pome}. It is understood that 
the cross sections $\sigma_{e,x}$ must be corrected for energy threshold 
and resolution effects, as described in the Appendix.

\subsection{Standard (no oscillation) case}

	In the standard (no oscillation) case, characterized by  $P(E,t)=1$, 
the signal $S$ varies as $S^{\rm std}(t) \propto L^{-2}(t)$. The standard 
Fourier components are simply given by 
\begin{equation}
f_n^{\rm std} = \varepsilon\delta_{n1}\ ,
\end{equation}
i.e., only the first harmonic is nonzero and measures the Earth's orbit
eccentricity.

\subsection{$2\,\nu$ oscillation case}

	In this case, $\nu_e$ is a linear combination of two mass eigenstates 
$(\nu_1,\,\nu_2)$ characterized by a mixing angle $\theta$
\begin{equation}
\nu_e=\cos\theta\, \nu_1 +\sin\theta\, \nu_2
\label{theta}
\end{equation}
and by a squared mass difference $\delta m^2$, 
\begin{equation}
\delta m^2 = |m^2_2 - m^2_1|  \ .
\label{dm2}
\end{equation}
The corresponding $\nu_e$ survival probability is given by
\begin{equation}
P^{2\nu}(E,t) = 1-\frac{1}{2}\sin^2 2\theta \left(1-\cos
\frac{\delta m^2\,L(t)}{2\,E} \right)\ .
\label{P2nu}
\end{equation}

	The Fourier components can be cast in  the following, compact form 
\cite{Four}:
\begin{equation}
f_n^{\rm 2\nu} = \frac{\varepsilon\delta_{n1}-\sin^2{2\theta}
			\,D_n(\delta m^2)}{1-\sin^2{2\theta}\,D_0
			(\delta m^2)}
\label{f2nu}
\end{equation}
where the (detector-dependent) functions $D_n$ are given by
\begin{equation}
	D_n(\delta m^2)=
	\frac{\displaystyle\int\!dE\,\lambda\,(\sigma_e-\sigma_x)\,U_n}
	{\displaystyle2\int\!dE\,\lambda\,\sigma_e}\;\;\;(n\geq 0)
\label{Dn}
\end{equation}
and the universal (i.e., detector-independent) functions $U_n$ are
given by
$$
	U_n(z)	=	\delta_{n0}-u_n(z)-\varepsilon
			[u_{n+1}(z)+u_{n-1}(z)-\delta_{n1}]\ ,
$$
$$
	u_n(z)	=	\cos\left(z-\frac{n\pi}{2}\right)J_n(\varepsilon z)\ ,
$$
where $z=\delta m^2 L_0/2 E$ and $J_n$ is the Bessel function of order $n$.
Notice that, although our calculations are of ${\cal O}( \varepsilon$), all 
orders in $\varepsilon z$ are kept, since $z$ may be large.

\subsection{$3\,\nu$ oscillation case}

	In this case,  $\nu_e$ is a linear combination of three mass 
eigenstates  $(\nu_1,\,\nu_2,\,\nu_3)$, usually parametrized in one of the 
following two forms:
\begin{eqnarray}
\nu_e &=& U_{e1}\,\nu_1+U_{e2}\,\nu_2+U_{e3}\,\nu_3\\
      &=& c_\phi(c_\omega\,\nu_1+s_\omega\,\nu_2)+s_\phi\,\nu_3\ ,
\label{nue}
\end{eqnarray}
where $s=\sin$, $c=\cos$, and $U^2_{e1}+U^2_{e2}+U^2_{e3}=1$
(see, e.g., \cite{Faid}). In addition, we assume a hierarchy of mass 
differences:
\begin{equation}
\delta m^2 \ll |m^2_3 - m^2_{1,2}|\ .
\label{m2}
\end{equation}
The hypothesis~(\ref{m2})  covers most of the situations of phenomenological 
interest \cite{Li97}. In this case, the $3\nu$ oscillation probability is 
given by
\begin{eqnarray}
P^{3\nu}(E,t) &=& 1-2\,U^2_{e3}(1-U^2_{e3})-2\,U^2_{e1}U^2_{e2}\left(1-\cos
\frac{\delta m^2\,L(t)}{2\,E} \right)\\
&=& c^4_\phi\, P^{2\nu}(E,t)\Big|_{\theta \to \omega} + s^4_\phi\ .
\label{P3nu}
\end{eqnarray}

	We have worked out the corresponding Fourier components, which read:
\begin{eqnarray}
f_n^{\rm 3\nu}&=& \frac{
		\varepsilon\,\delta_{n1}[1-4\, U^2_{e3}(1-U^2_{e3})\,r]
		-4\, U^2_{e1} U^2_{e2}\,D_n(\delta m^2)}
		{1-4\, U^2_{e3}(1-U^2_{e3})\,r-
		4\,U^2_{e1} U^2_{e2}\,D_0(\delta m^2)}\\
	&=& \frac{\varepsilon\,\delta_{n1}(1-s^2_{2\phi}r)
	-s^2_{2\omega}c^4_\phi\,D_n(\delta m^2)}
	{1-s^2_{2\phi}r-s^2_{2\omega}c^4_\phi D_0(\delta m^2)}
\label{f3nu}
\end{eqnarray}
where the functions $D_n$ are defined in Eq.~(\ref{Dn}) and
\begin{equation}
r=\frac{\displaystyle\int\! dE\, \lambda(\sigma_e-\sigma_x)}
{\displaystyle 2\int\! dE\,\lambda\sigma_e }\ .
\end{equation}
Notice that $f^{3\nu}\to f^{2\nu}$ for $U_{e3}(=s_\phi)\to 0$, as it should.

\subsection{A useful consistency check}

	{\em A priori}, the signal $S(t)$ must obey the symmetry $S(t)=S(T-t)$ 
in the one-year interval $[0,\,T]$  (either with 
or without oscillations). We have 
made use of this property in Eqs.~(\ref{S}--\ref{fnexpt}). The experimental 
test of such symmetry property is not without merit, since its failure 
might signal systematic, time-dependent effects, such as unexpected 
variations of the detection efficiency or of the background level.  
Within our approach, this is equivalent to check that the ``sine'' Fourier 
components $g_n$,  defined as
\begin{equation}
	g_n 	 =  	\frac{1}{N_S}\sum_{i=1}^{N_S+N_B}
 			\sin\frac{2\pi n t_i}{T}\ \ \ \ \ \
 			(0\leq t_i\leq T)\ ,
\label{gnexpt}
\end{equation}
are identically zero, as they should (both in the  standard case and in the 
presence of possible oscillations):
\begin{equation}
g_n^{\rm std} = g_n^{\rm osc}= 0\ . 
\end{equation}
Notice that, in Eq.~(\ref{gnexpt}), the event ``arrival times'' are folded
in the interval $[0,\,T]$ and not in  $[0,\,T/2]$.

	The experimentally inferred $g_n$'s will be distributed around zero 
with a variance ${\rm var} (g_n)$.  As far as statistical fluctuations are 
concerned, the  calculation of ${\rm var} (g_n)$ is analogous to ${\rm var} 
(f_n)$ \cite{Four} and gives:
\begin{equation}
	{\rm var}(g_n)	= \frac{1-f_{2n}+N_B/N_S}{2\,N_S} \ .
\label{vargn}
\end{equation}
However, as already noticed, the $f_{2n}$'s are generally $\ll 1$ and 
thus can be neglected in the above equation. Therefore, the statistical
uncertainty $\sigma_g=\sqrt{{\rm var}(g_n)}$ is approximately equal for all 
the $g_n$'s and  has the same expression as $\sigma_f$ [Eq.~(\ref{sigma})].

	In conclusion, it is useful to check that the semiannual variations 
of the solar neutrino signal are indeed symmetric in time.  This implies that 
the ``sine'' Fourier components $g_n$ defined in Eq.~(\ref{gnexpt}) should
form a distribution with mean value $\langle g_n \rangle=0$  and standard 
deviation $\sigma_g=\sqrt{(N_S+N_B)/2 N_S^2}$. This test is universal, i.e.\ 
it is valid with or without oscillation effects. Any deviation of the mean 
and variance of the $g_n$ distribution from their standard values (0 and 
$\sigma^2_g$, respectively) would indicate the presence (and the magnitude) 
of systematic effects beyond the purely  statistical fluctuations of 
either the background or the  signal. These effects (if any)  should then be 
accounted for by the experimentalists, before performing  an unbiased analysis 
of the time variations of the signal.

\section{Vacuum oscillation tests at SuperKamiokande and Borexino}

	In this section we analyze the tests of the vacuum oscillation 
hypothesis that can be performed at SuperKamiokande and Borexino, both 
separately and jointly. We consider two observables for each experiment.
In particular, we analyze $\Delta \langle T \rangle/\langle T \rangle$ and 
$f_1-\varepsilon$ for SuperKamiokande (SK), and $f_1-\varepsilon$ 
and $f_2$ for Borexino (BX):
\begin{equation}
{\rm Observables\ }\left\{ 
\begin{array}{l}
\Delta \langle T \rangle/\langle T \rangle
 {\rm\  and\ } f_1^{\rm SK}-\varepsilon {\rm\ \ (SuperKamiokande)}\ ,\\
 f_1^{\rm BX}-\varepsilon {\rm\  and\ } f_2^{\rm BX} {\rm\ \ (Borexino)}\ .
\end{array}
\right.
\end{equation}
All the above variables are zero in  the standard (no oscillation) case.
In the oscillation range of interest, Fourier components with $n >1 $ 
($n> 2$) are not relevant for SuperKamiokande (Borexino) \cite{Four}.
For simplicity, we will consider only $2\nu$ oscillations, and the 
corresponding preferred regions A, B, C, and D of Fig.~1.

\subsection{Tests at SuperKamiokande}

	Figure~3 shows the four solutions A, B, C, and D (gray regions)
and the no oscillation point (star) in the plane charted by  the parameters  
$\Delta \langle T \rangle/\langle T \rangle$ (the fractional deviation
of the mean electron kinetic energy) and $f_1^{\rm SK}-\varepsilon$ (the 
deviation of the first Fourier component from its standard value). Also 
shown is the horizontal band allowed at $\pm 1\sigma$ by the 
$\Delta \langle T \rangle/\langle T \rangle$ datum of Eq.~(\ref{DT}).

	Figure~3 evidences that solutions C and D (which predict large, 
negative values for  $\Delta \langle T \rangle/\langle T \rangle$) are 
highly disfavored by the SuperKamiokande measurement of Eq.~(\ref{DT}). In 
particular, solution  C is disfavored at $>4\sigma$ and solution D at 
$>6\sigma$. On the other hand, the datum of Eq.~(\ref{DT}) is unable to 
discriminate among the solutions A, B, and the no oscillation point at the 
$2\sigma$ level, although solution A seems to be preferred. We can 
summarize these findings by saying that, under the hypothesis of $2\nu$ 
oscillations, the {\em combined}  data of Table~I and Eq.~(\ref{DT})
select the solutions A and B in Fig.~1, corresponding to the following 
approximate ranges (at 95\% C.L.)  for the neutrino mass-mixing parameters:
\begin{eqnarray}
& 0.59 \lesssim \delta m^2\lesssim  0.84\ (\times 10^{-10}{\rm\ eV}^2)&\ ,\\
& 0.66 \lesssim \sin^2 2\theta \lesssim 1 &\ .
\end{eqnarray}
Notice that the spread of the above parameters is only about $\pm 20\%$.
Also notice that $2\nu$ maximal mixing $(\sin^2 2\theta = 1)$ is 
allowed only in a restricted range of $\delta m^2$ 
($\sim 0.59$--$0.61\times 10^{-10}$ eV$^2$).

	Concerning $f_1$, we cannot infer its value from the limited
data which are publicly available \cite{To97,Na98,In97,Su97,It97,So98}. 
At any rate,   the estimated uncertainty  $\sigma_f$ of $f_1$ 
[see Eq.~(\ref{sigma})] for $\sim 1$~yr of data  taking is about 
$\pm 0.02$  \cite{Fo97}---too large to discriminate any of the 
solutions A, B, C, and D. Significantly 
higher statistics are needed to reduce such error.

	An interesting feature of Fig.~3 is the tight correlation between
the variables $f_1-\varepsilon$ and $\Delta \langle T\rangle/\langle T\rangle$, 
which parametrize $L$-related and $E_\nu$-related spectral distortions, 
respectively. Such correlation can 
be traced to the $L/E_\nu$ dependence of the oscillation probability, as 
emphasized in \cite{Mi97}. One can use such correlation
to ``predict'' the value of $f_1-\varepsilon$ in SuperKamiokande for a given
value of $\Delta \langle T \rangle/\langle T \rangle$ in Eq.~(\ref{DT}). 
More precisely, from Fig.~3 one derives that the values of $f_1-\varepsilon$ 
compatible with both the horizontal band and the solution A should lie in the 
range $\sim 0.005$--$0.01$. In other words, one expects an enhancement 
of the semiannual modulations of the neutrino flux (relatively to the purely 
geometrical one) in the range 
$[0.005/\varepsilon,\,0.01/\varepsilon]=30$--$60\%$ at $\sim 1\sigma$.
This is a clear prediction that needs, however, several years of data 
taking at SuperKamiokande to be tested.

\subsection{Tests at Borexino}

	Figure~4 shows the four solutions A, B, C, and D (gray regions)
in the plane charted by the parameters $f_1^{\rm BX}-\varepsilon$ and 
$f_2^{\rm BX}$ (the deviation of the first two Fourier components from 
their standard values). The  no oscillation point (star) corresponds 
to the origin.

	The amplitude of the second harmonic appears to be generally 
smaller than the first; nevertheless, both should be detectable in a sample 
of a few thousand events. E.g., for $N_S \simeq N_B \simeq 5000$ events, 
the expected statistical error of $f_1$ and $f_2$ is only about $\pm 0.014$ 
[see Eq.~(\ref{sigma})], ensuring clear detection of semiannual modulations 
(provided that systematics do not dominate the error budget). However, the 
four solutions A, B, C, D are rather close to each other in the Fourier 
parameter space, and it might be difficult to distinguish among them using 
only these two variables.

	By comparing Figs.~3 and 4, it can be noticed that the solutions
A and B (currently favored by SuperKamiokande data) predict rather 
different values for the first Fourier harmonic in SuperKamiokande and 
Borexino, as a result of the different energy ranges probed by these two
experiments. In particular,  within solution A it is always 
$f_1^{\rm SK}-\varepsilon>0$, while $f_1^{\rm BX}-\varepsilon$ can be either 
positive or negative.  Therefore, at present the sign of the semiannual 
variations in Borexino is unpredictable.

	It is interesting to notice that, both in $2\nu$ and $3\nu$ 
oscillations, the Fourier component ratio $f_2/(f_1-\varepsilon)$ depends 
only on $\delta m^2$  and not on the mixing angle(s) [see Eqs.~(\ref{f2nu}) 
and (\ref{f3nu})].  Therefore, such ratio can constrain the value of 
$\delta m^2$ in a model-independent way.

\subsection{Combination of SuperKamiokande and Borexino tests}

	Figure~5 shows the four solutions A, B, C, and D (gray regions)
in the four planes charted by  the Borexino observables 
$f_1^{\rm BX}-\varepsilon$ and $f_2^{\rm BX}$ ($x$-axes) and by the 
SuperKamiokande observables $f_1^{\rm SK}-\varepsilon$  and 
$\Delta \langle T \rangle /\langle T \rangle$ ($y$-axes). Also shown 
is the horizontal band allowed at $\pm 1\sigma$ by the 
$\Delta \langle T \rangle/\langle T \rangle$ datum of Eq.~(\ref{DT}).
Within such band, the values of $f_1^{\rm BX}-\varepsilon$ 
(as well as those of
$f_2^{\rm BX}$) can be either positive or negative, as observed
in the previous subsection. Figure~5 shows that the present indeterminacy 
in the sign of  $f_1^{\rm BX}-\varepsilon$ cannot be resolved by 
increasing the accuracy of the SuperKamiokande data.

	The similarity between the upper and lower panels in Fig.~5 is due to
the tight correlation between the SuperKamiokande variables  
$f_1^{\rm SK}-\varepsilon$  and $\Delta \langle T \rangle /\langle T \rangle$. 
This similarity should be reflected in the experimental data, if vacuum 
oscillations indeed occur. When experimental data will be available for 
all the four observables charting the panels in Fig.~5, one of 
the four solutions should be easily spotted in at least one of the four
panels. However, in the unlucky situation of data points close to
the no oscillation case, it might be difficult to distinguish between 
such case and solution B. The possibility to separate the no oscillation 
point from solution B would then depend decisively on the reduction of 
the $\Delta \langle T \rangle /\langle T \rangle$ uncertainty (see upper 
panels of Fig.~5). In any case, the reader can judge the discriminating 
power of the two experiments  by drawing prospective data points and error 
bars in each panel of Fig.~5.

\section{Summary  and Conclusions}

	We have studied tests of vacuum oscillations of solar neutrinos
in the SuperKamiokande and Borexino experiments, both separately and 
jointly. The tests are sensitive to either energy or time variations of 
the neutrino flux. The interplay between such tests has been investigated, 
in the light of the most recent data (374 day) from the SuperKamiokande 
experiment. The results have been  displayed in a graphical form 
(Figs.~3--5) that allows to determine easily the experimental accuracy needed 
to test the vacuum oscillation solution(s). We have found that: 
(i) The total neutrino rates measured by solar $\nu$ experiments can be 
fitted in four distinct regions of the mass-mixing parameter space;
($ii$) Two of the four solutions are strongly disfavored by the 
SuperKamiokande energy spectrum data; $(iii)$ The energy spectrum data
do not discriminate significantly (at present) the remaining two solutions
between them and from the no oscillation case; ($iv)$ The amplitude of
semiannual variations of the solar $\nu$ flux in SuperKamiokande 
is predicted to be about 30--60\% in excess of the purely geometric one
(at $1\sigma$); $(v)$ The sign of semiannual variations (due to oscillations)
in Borexino is not determined by present data; $(vi)$ The {\em joint}
information coming from the energy spectrum data (SuperKamiokande) 
and from the Fourier transform  of the solar neutrino rates 
(SuperKamiokande, Borexino) can provide powerful tests of the 
vacuum oscillation solutions. Besides,
we have generalized the Fourier transform formalism  to three-flavor 
oscillations, and we have discussed a useful check  of the time 
symmetry  of the signal.

\acknowledgements

	We are grateful to J.\ G.\ Learned for useful discussions and helpful
suggestions. EL thanks the Centro de F{\'\i}sica das Interac\c{c}\~oes
Fundamentais at the  Instituto Superior T{\'e}cnico of the University of 
Lisbon, Portugal, for kind hospitality during the preparation of this work.
BF thanks the Dipartimento di Fisica of Bari, Italy, for hospitality.

\appendix
\section*{Energy threshold and resolution effects}

	In the calculation of the expected signal [Eq.~(\ref{Sosc})], it is
understood that the $\nu_\alpha$-$e$ cross sections $\sigma_\alpha(E)$ 
$(\alpha = e,\,x)$ have to be properly corrected to take into account the
detector energy resolution and the analysis window for each experiment.
Here we give some details about such corrections.

	Both in Borexino and in SuperKamiokande, the finite energy resolution 
due to the photon statistics implies that the {\em measured\/} kinetic energy
$T$ of the scattered electron is distributed around the {\em true\/} 
kinetic energy $T'$ according to a resolution function $R(T,\,T')$ of
the form \cite{Kr97}:
\begin{equation}
R(T,\,T') = \frac{1}{\sqrt{2\pi}s}\exp\left[
{-\frac{(T-T')^2}{2 s^2}}\right]\ ,
\end{equation}
where
\begin{equation}
s = s_0\sqrt{T'/{\rm MeV}}\ ,
\label{Delta}
\end{equation}
and $s_0=57.7$ KeV and 0.47 MeV for Borexino \cite{Be95}
and SuperKamiokande \cite{To97,Fo97}, respectively. On the other hand, the 
distribution of the true kinetic energy $T'$ for an interacting neutrino of 
energy $E_\nu$ is dictated by the differential cross section 
$d\sigma_\alpha(E_\nu,\,T')/dT'$, that we take from \cite{CrSe}. The 
kinematic limits are:
\begin{equation}
0\leq T' \leq {\overline T}'(E_\nu)\ , 
\ \ {\overline T}'(E_\nu)=\frac{E_\nu}{1+m_e/2E_\nu}\ .
\end{equation}

	Concerning the measured kinetic energy, the present analysis window 
for SuperKamiokande is 
$[T_{\rm min},\,T_{\rm max}]=[6.5{\rm\ MeV}-m_e,\,20{\rm\ MeV}-m_e]$.
The prospective analysis window for Borexino (as used in this paper)
is $[0.25,\,0.8]$ MeV.

	For assigned values of $s_0$, $T_{\rm min}$, and $T_{\rm max}$, the
corrected cross section $\sigma_\alpha(E_\nu)$ is defined as:
\begin{equation}
\sigma_\alpha(E_\nu)=\int_{T_{\rm min}}^{T_{\rm max}}\!dT
\int_0^{{\overline T}'(E_\nu)}
\!dT'\,R(T,\,T')\,\frac{d\sigma_\alpha(E_\nu,\,T')}{dT'}\ .
\end{equation}
It is useful to reorder the integrands, obtaining
\begin{equation}
\sigma_\alpha(E_\nu)=
\int_0^{{\overline T}'(E_\nu)}
\!dT'\,W(T')\,\frac{d\sigma_\alpha(E_\nu,\,T')}{dT'}\ ,
\end{equation}
where the function $W(T')$, which embeds the detector specifications
$s_0$, $T_{\rm min}$, and $T_{\rm max}$, is given by
\begin{equation}
W(T') = \frac{1}{2}\left[ 
{\rm erf}\left(\frac{T_{\rm max}-T'}{\sqrt{2}\,s } \right) - 
{\rm erf}\left(\frac{T_{\rm min}-T'}{\sqrt{2}\,s } \right)
\right]\ ,
\end{equation}
with $s$ as in Eq.~(\ref{Delta}) and
\begin{equation}
{\rm erf}(x)=\frac{2}{\sqrt{\pi}}\int_0^x\!dt\,e^{-t^2}\ .
\end{equation}


\begin{table}
\caption{Neutrino event rates measured by solar neutrino experiments,
	and corresponding predictions from the  standard solar model
	\protect\cite{BP95}.
	The quoted errors are at $1\sigma$.}
\begin{tabular}{ccccc}
Experiment 	&Ref.& Data~$\pm$(stat.)~$\pm$(syst.)
& Theory \protect\cite{BP95}& Units \\
\tableline
Homestake	& \protect\cite{La97}& $ 2.56 \pm 0.16 \pm 0.15$         &
	$9.3^{+1.2}_{-1.4}$    & SNU					\\
Kamiokande 	& \protect\cite{Fu96}& $ 2.80 \pm 0.19 \pm 0.33$         &
	$6.62^{+0.93}_{-1.12}$ & $10^6$ cm$^{-2}$s$^{-1}$		\\
SAGE		& \protect\cite{Ga97}&$69.9^{+8.5}_{-7.7}{}^{+3.9}_{-4.1}$&
	$137^{+8}_{-7}$        & SNU					\\
GALLEX		& \protect\cite{Hp97}& $76.4 \pm  6.3^{+4.5}_{-4.9}$     &
	$137^{+8}_{-7}$	       & SNU					\\
SuperKam.	& \protect\cite{To97}&$2.37^{+0.06}_{-0.05}{}^{+0.09}_{-0.06}$&
 	$6.62^{+0.93}_{-1.12}$ & $10^6$ cm$^{-2}$s$^{-1}$	
\end{tabular}
\end{table}




\begin{figure}
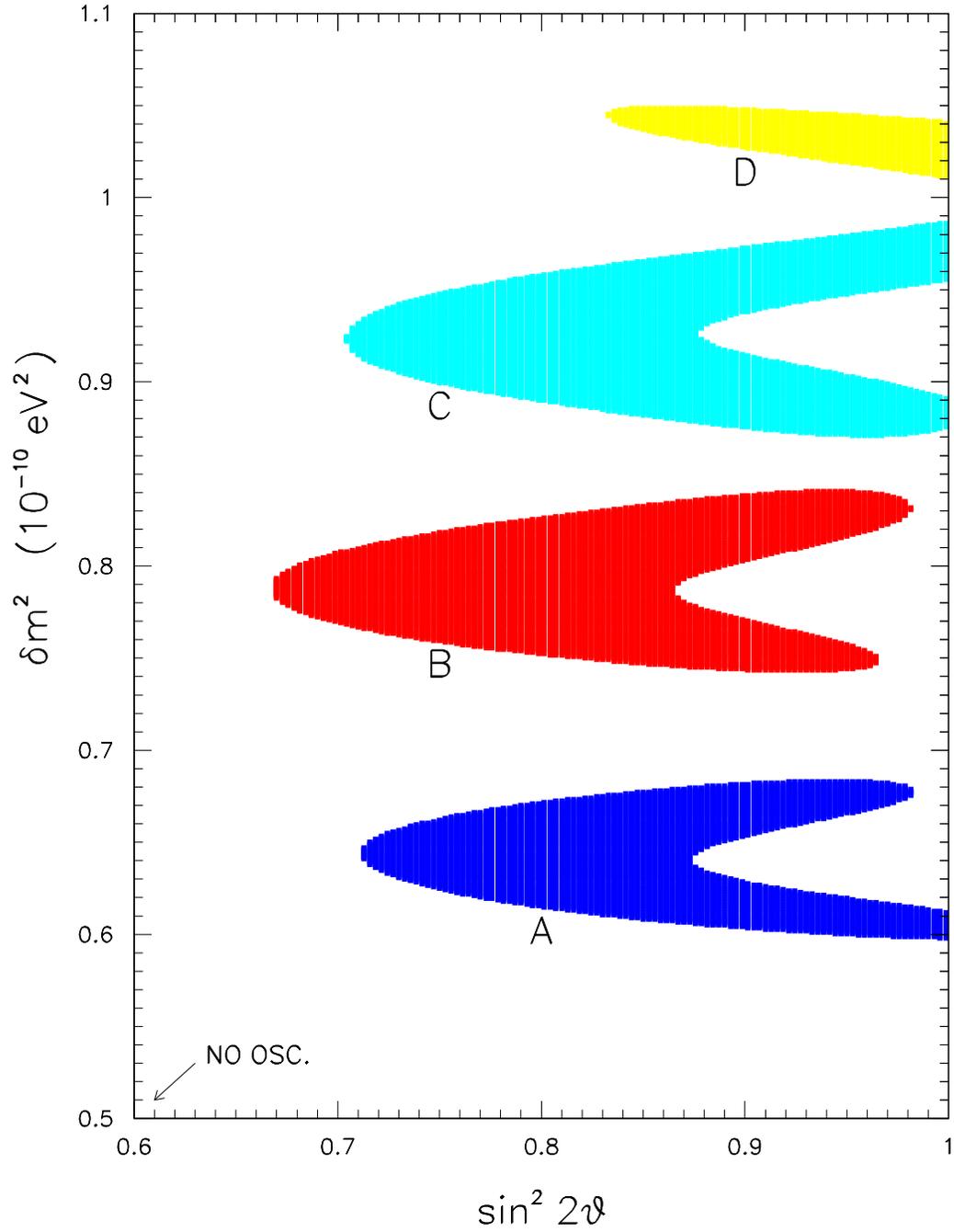

\caption{Vacuum oscillation solutions to the solar neutrino deficit
	in the usual mass-mixing plane, as derived from a fit to the 
	data of Table~I. The four regions A, B, C, and D, are allowed
	at 95\% C.L.}
\label{fig:1}
\end{figure}
\begin{figure}
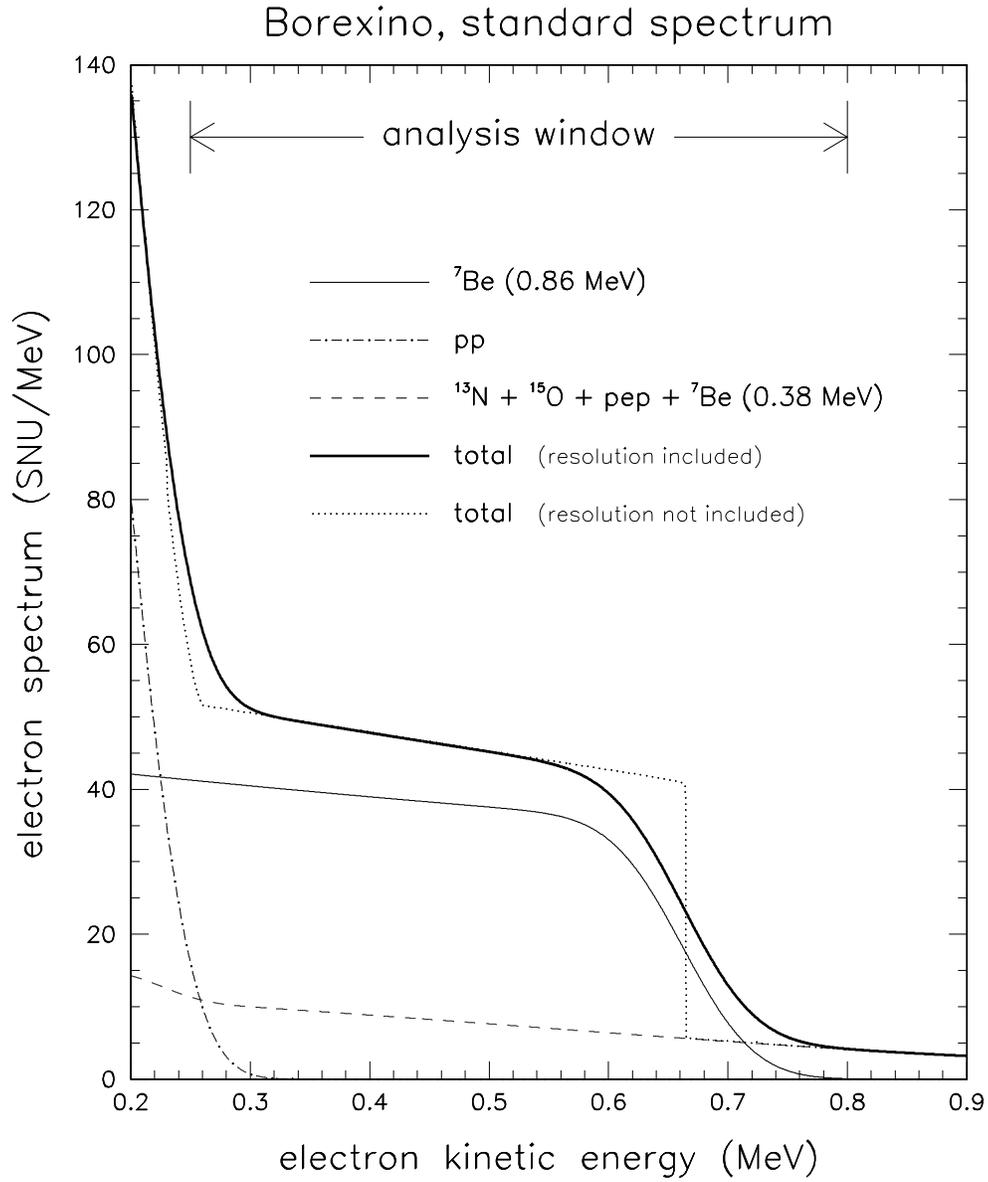

\caption{Energy spectrum of recoil electrons in Borexino. The main components
	are shown separately. The smearing effect of the energy resolution
	is also shown. The arrows indicate the prospective energy window 
	assumed in the analysis.}
\label{fig:2}
\end{figure}
\begin{figure}
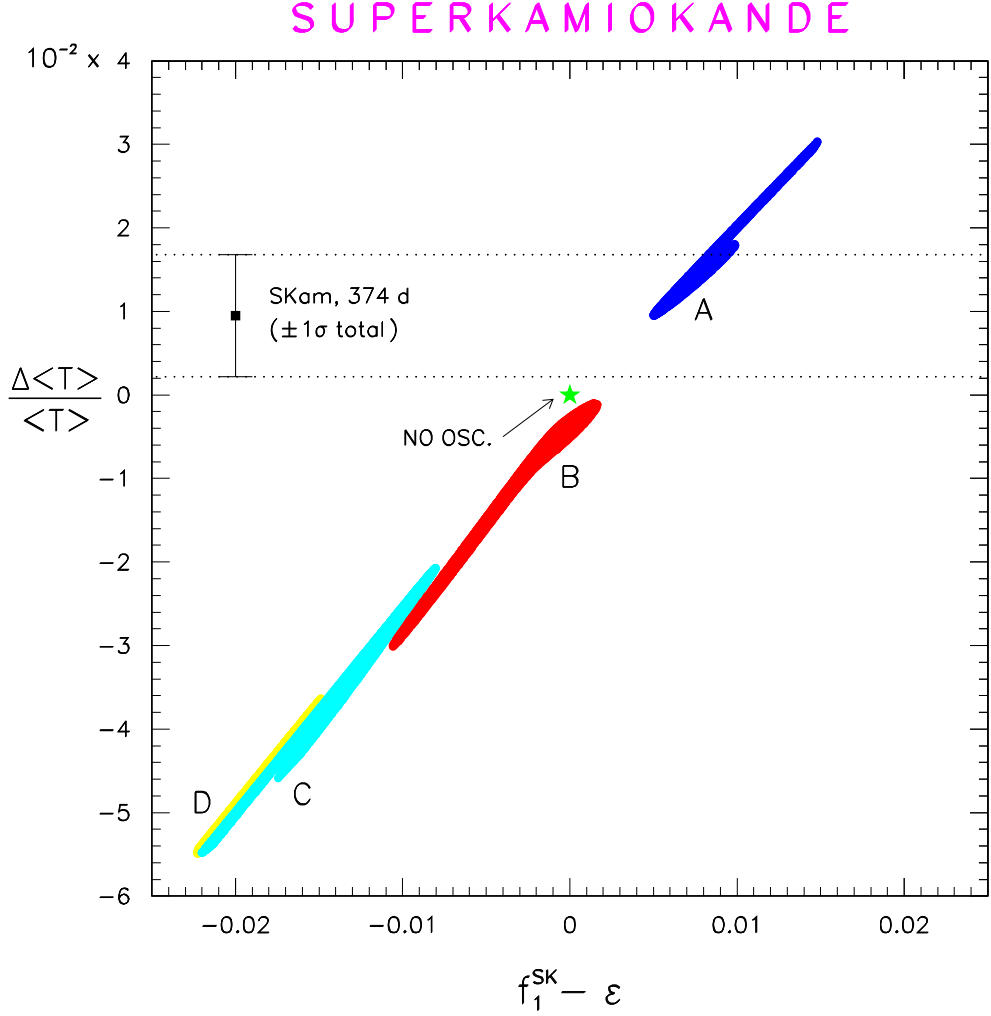

\caption{SuperKamiokande (SK) experiment. 
	Map of the four vacuum oscillation solutions A, B, C, and D 
	in the plane charted by $\Delta \langle T \rangle/\langle T \rangle$
	(the fractional deviation of the mean electron kinetic energy)
	and by $f_1-\varepsilon$ (the deviation of the first Fourier
	component from its standard value).
	Notice the strong correlation between these two variables, which
	is induced by the $L/E$ dependence of the oscillation probability.
	The star at the origin corresponds to
	the standard (no oscillation) case.
	The SuperKamiokande datum on 
	$\Delta \langle T \rangle/\langle T \rangle$
	is also shown; it disfavors solution C at $>4\sigma$ and
	solution D at $>6 \sigma$. The part of the solution $A$ which
	is favored by the SK datum at $1\sigma$ corresponds to an
	expected Fourier amplitude $f_1-\varepsilon\simeq 0.5$--$1\%$.}
\label{fig:3}
\end{figure}
\begin{figure}
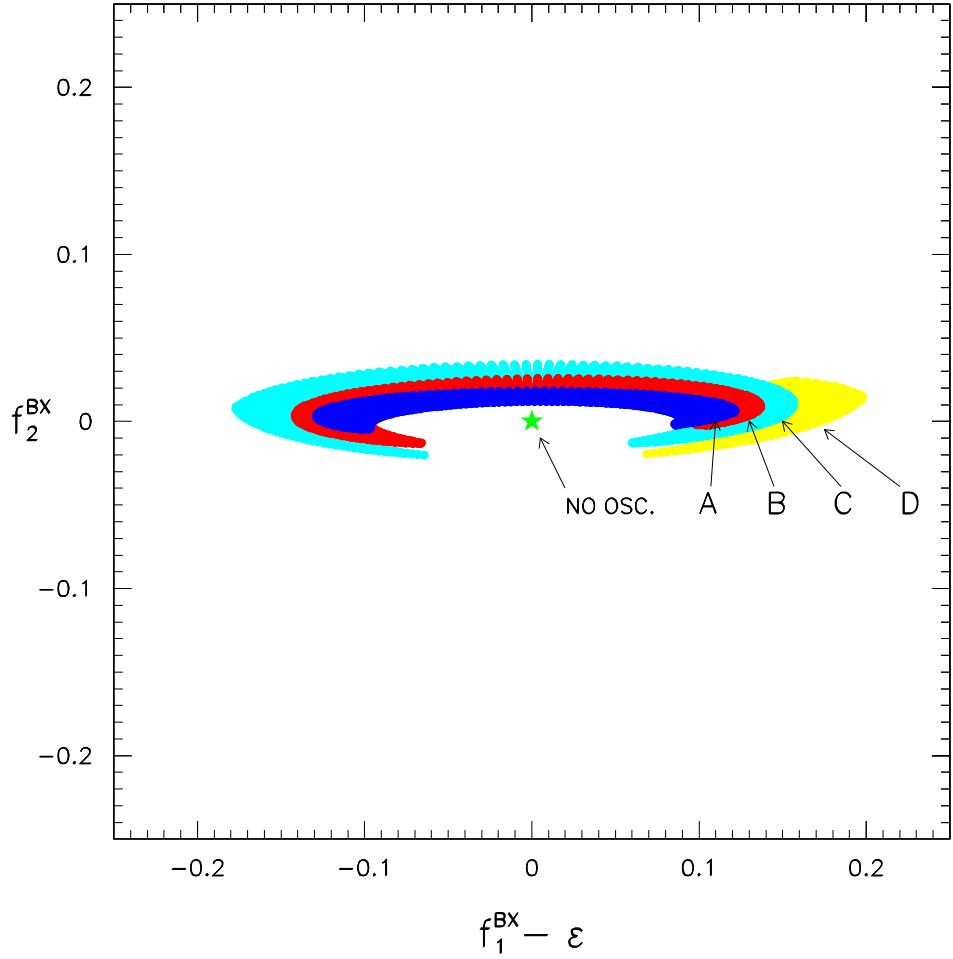

\caption{Borexino (BX) experiment. Map of the four vacuum oscillation solutions
	A, B, C, and D, in the plane charted by the first two Fourier
	components. The star at the origin corresponds to the no oscillation
	case. There is some overlap among the four solutions.}
\label{fig:4}
\end{figure}
\begin{figure}
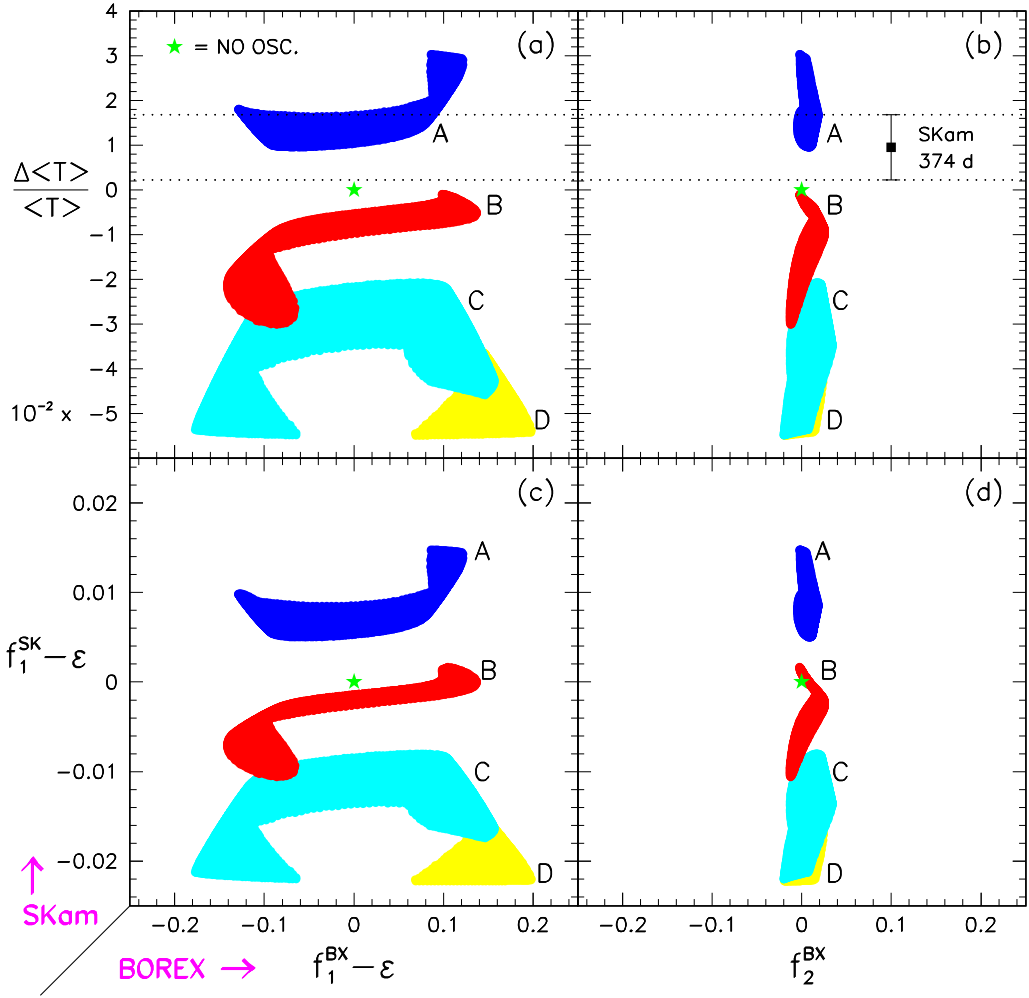

\caption{SuperKamiokande (SK) {\em vs\/} Borexino (BX). Map of the four
	vacuum oscillation solutions A, B, C, and D in the planes charted
	by the SK observables $\Delta \langle T \rangle/\langle T \rangle$
	and $f_1-\varepsilon$ (ordinates) and by the BX observables
	$f_1-\varepsilon$ and $f_2$ (abscissae). Future measurements of the
	$f_n$'s in SK and BX, together with increasingly accurate
	$\Delta \langle T \rangle/\langle T \rangle$ data from SK, 
	are expected to spot one of the solutions.}
\label{fig:5}
\end{figure}



\newcommand{\InsertFigure}[2]{\newpage\begin{center}\mbox{%
\epsfig{bbllx=1.4truecm,bblly=1.3truecm,bburx=19.5truecm,bbury=26.5truecm,%
height=20.truecm,figure=#1}}\end{center}\vspace*{-1.7truecm}%
\parbox[t]{\hsize}{\small\baselineskip=0.45truecm\hskip0.5truecm #2}}

\InsertFigure{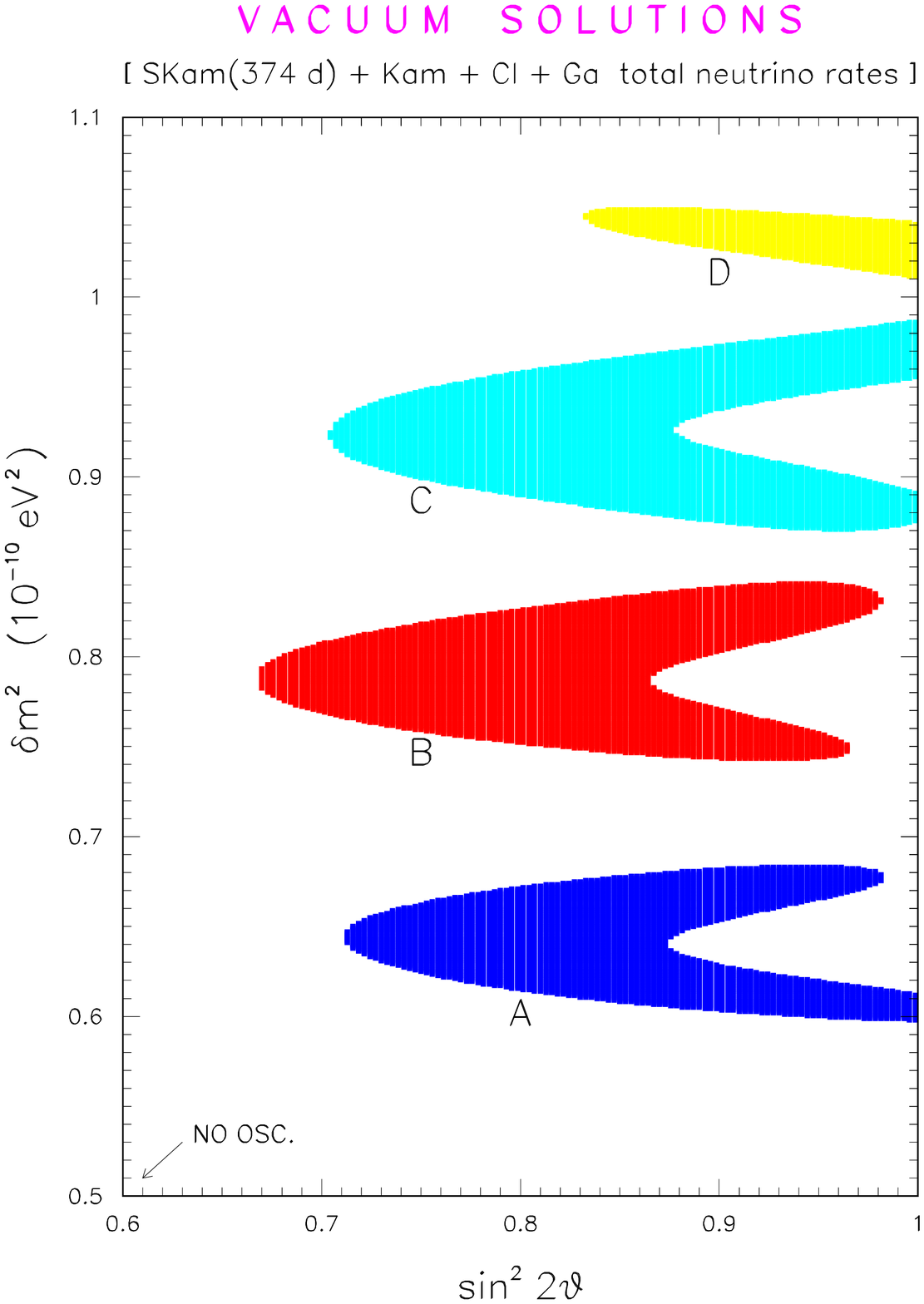}%
{FIG.~1.  Vacuum oscillation solutions to the solar neutrino deficit
	in the usual mass-mixing plane, as derived from a fit to the 
	data of Table~I. The four regions A, B, C, and D, are allowed
	at 95\% C.L.}
\InsertFigure{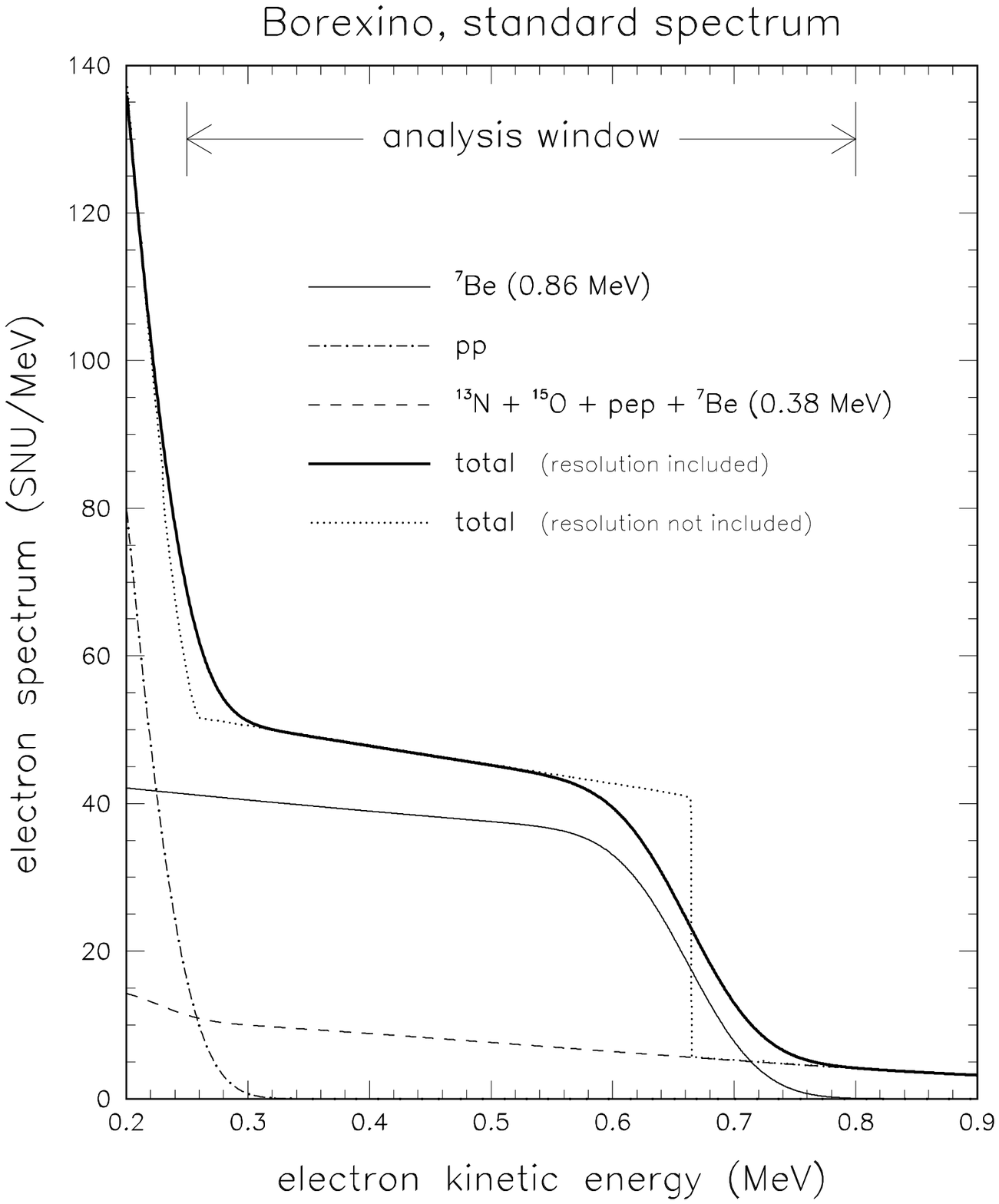}%
{FIG.~2. Energy spectrum of recoil electrons in Borexino. The main components
	are shown separately. The smearing effect of the energy resolution
	is also shown. The arrows indicate the prospective energy window 
	assumed in the analysis.}
\InsertFigure{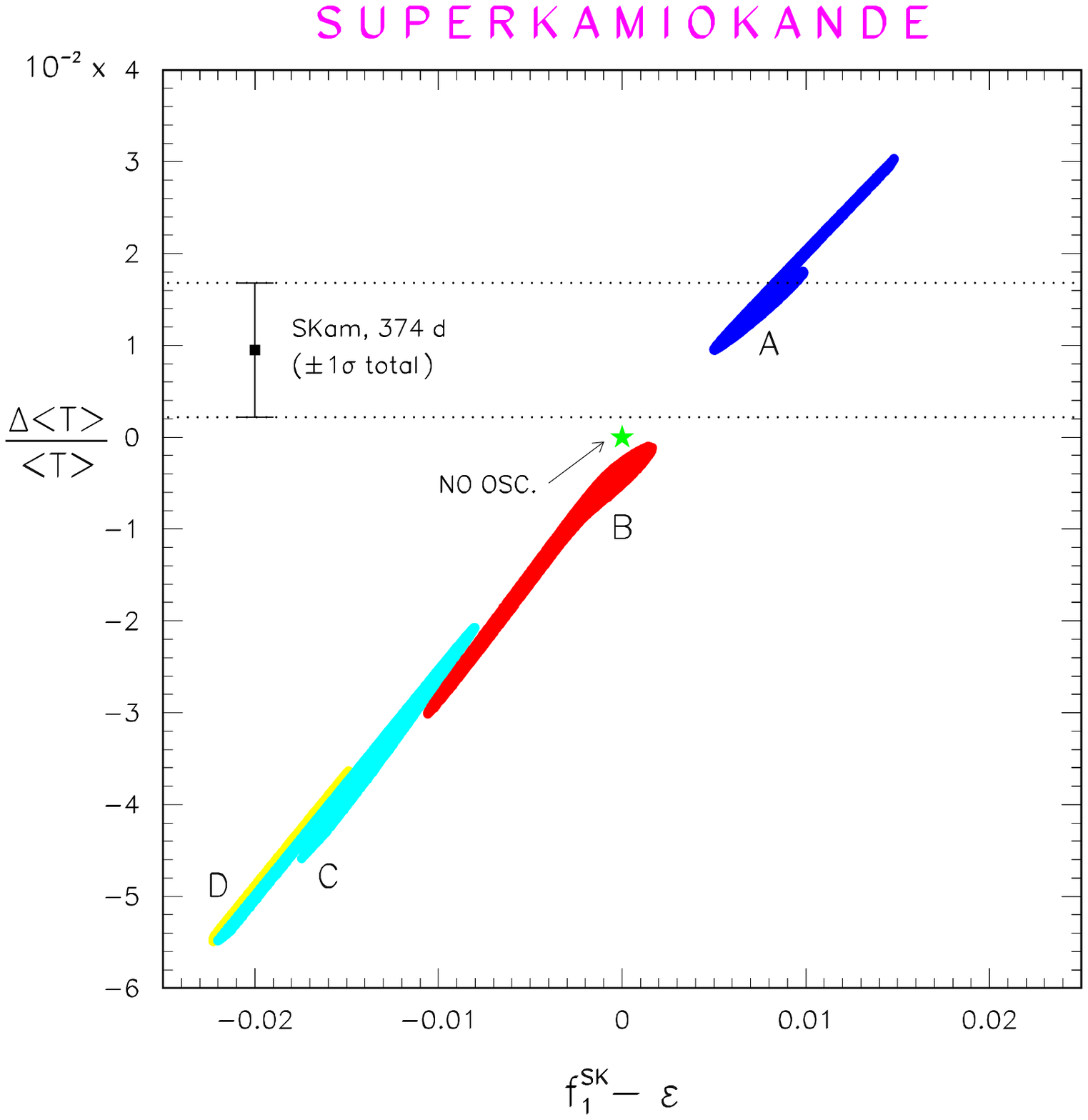}%
{FIG.~3.  SuperKamiokande (SK) experiment. 
	Map of the four vacuum oscillation solutions A, B, C, and D 
	in the plane charted by $\Delta \langle T \rangle/\langle T \rangle$
	(the fractional deviation of the mean electron kinetic energy)
	and by $f_1-\varepsilon$ (the deviation of the first Fourier
	component from its standard value).
	Notice the strong correlation between these two variables, which
	is induced by the $L/E$ dependence of the oscillation probability.
	The star at the origin corresponds to
	the standard (no oscillation) case.
	The SuperKamiokande datum on 
	$\Delta \langle T \rangle/\langle T \rangle$
	is also shown; it disfavors solution C at $>4\sigma$ and
	solution D at $>6 \sigma$. The part of the solution $A$ which
	is favored by the SK datum at $1\sigma$ corresponds to an
	expected Fourier amplitude $f_1-\varepsilon\simeq 0.5$--$1\%$.}	
\InsertFigure{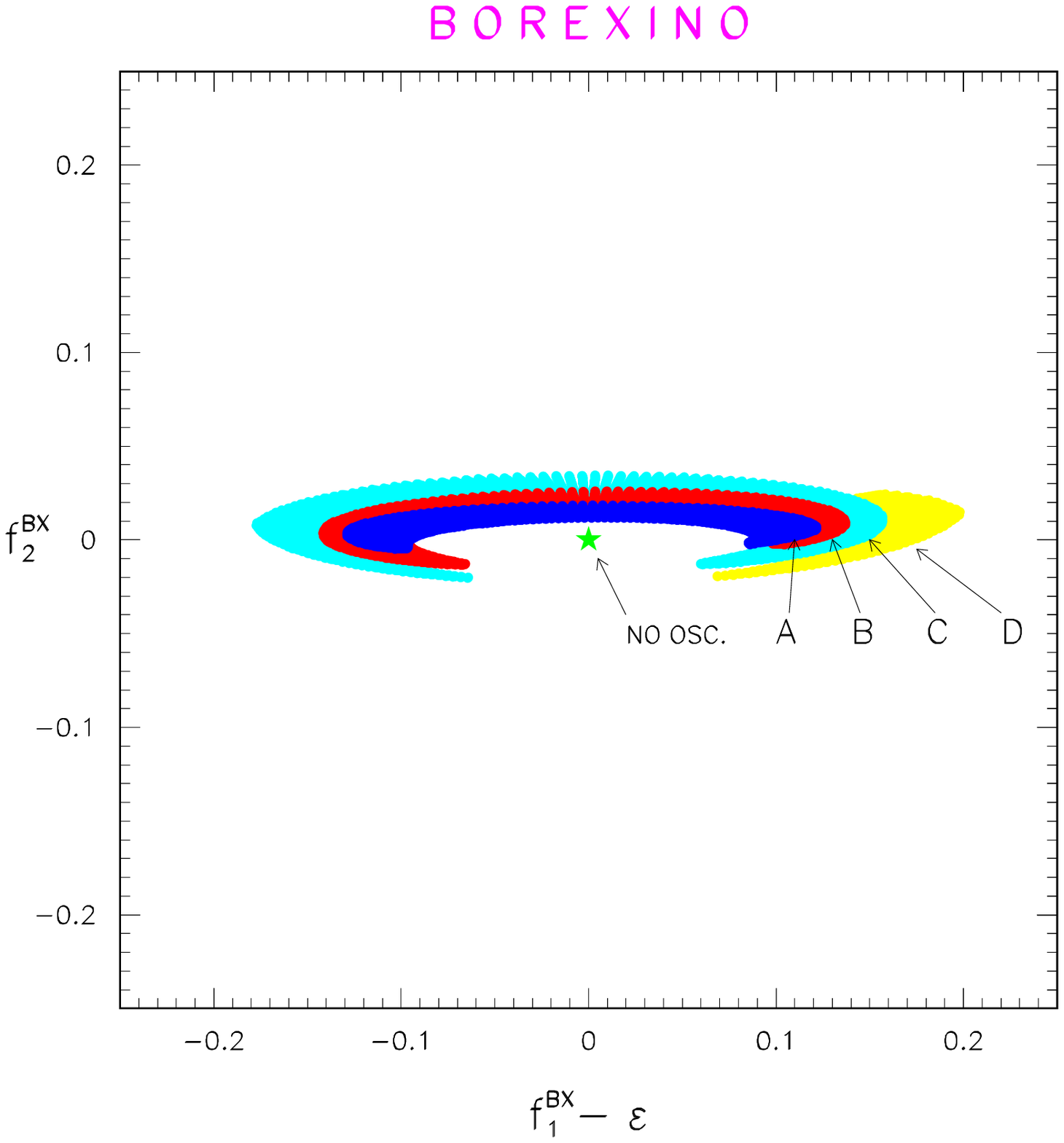}%
{FIG.~4.	Borexino (BX) experiment. Map of the four vacuum oscillation solutions
	A, B, C, and D, in the plane charted by the first two Fourier
	components. The star at the origin corresponds to the no oscillation
	case. There is some overlap among the four solutions.}
\InsertFigure{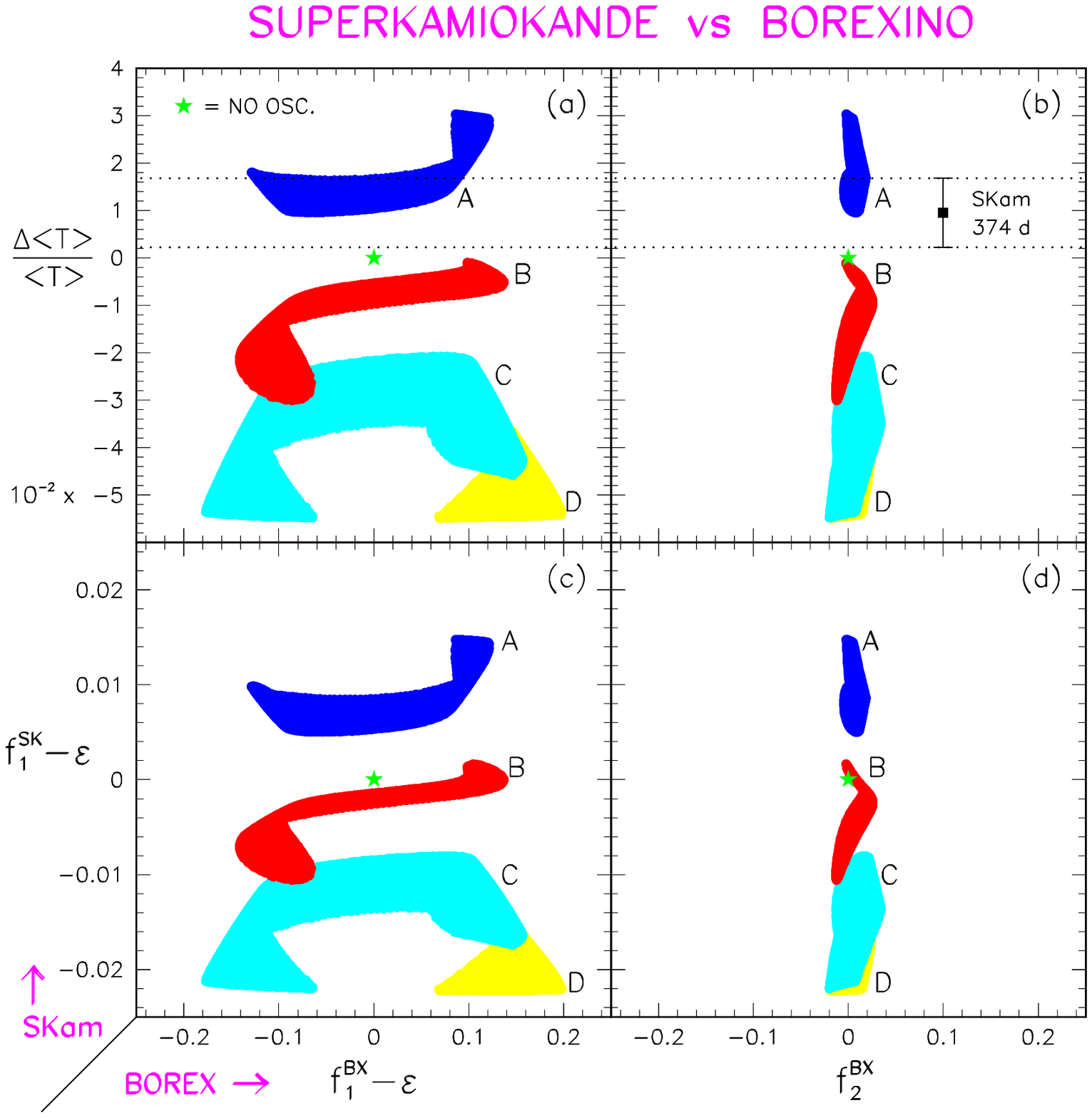}%
{FIG.~5.  SuperKamiokande (SK) {\em vs\/} Borexino (BX). Map of the four
	vacuum oscillation solutions A, B, C, and D in the planes charted
	by the SK observables $\Delta \langle T \rangle/\langle T \rangle$
	and $f_1-\varepsilon$ (ordinates) and by the BX observables
	$f_1-\varepsilon$ and $f_2$ (abscissae).
	Future measurements of the
	$f_n$'s in SK and BX, together with increasingly accurate
	$\Delta \langle T \rangle/\langle T \rangle$ data from SK,
	are expected to spot one of the solutions.}

\end{document}